\documentclass[aps,twocolumn,prl]{revtex4}
\usepackage{amsmath,amssymb}

\newcommand\av[1]{\langle {#1}\rangle }
\newcommand\gN[1]{\langle \!\langle g^{#1}\rangle \!\rangle }

\newcommand\gB[1]{\langle {g^{#1}}\rangle{\lower
4pt\hbox{$ \scriptstyle{\!B} $}}}
\newcommand\gBc[1]{\langle \!\langle {g^{#1}}\rangle\!\rangle {\lower
4pt\hbox{$ \scriptstyle{\!B} $}}}

\begin{document}


\noindent\textbf{Comment on ``Anomalous Conductance Distribution in Q1D Gold
Wires: Possible Violation of the One-Parameter Scaling Hypothesis''.}

Mohanty and Webb \cite{MW} claim that their data on conductance fluctuations
in gold wires contradict the one-parameter scaling. We show that flaws in
extracting values of the cumulants $\langle \!\langle g^{n }\rangle
\!\rangle $ of the conductance distribution (for $n\!=\!3,4$) invalidate all
the conclusions made there. The actual values of $\langle \!\langle g^{n
}\rangle \!\rangle $ determined by us from the published raw data contained
in Ref.\ \cite{MW} are orders of magnitude smaller than those claimed. We
show that a visible (but small) deviation of the distribution in \cite{MW}
from the normal shape results from a systematic error due to the limited
applicability of the ergodicity hypothesis. Thus, the data of \cite{MW} do
not warrant any statement on the violation or validity of the one-parameter
scaling.

1.\ {\bfseries\itshape {Comparing data with theory.}} Table I below copies
part of the Table in \cite{MW} as cumulants extracted from
magnetofingerprints $\delta g(B)$ (in units of $e^{2}/h$) measured in 3
samples. Comparing \cite{MW} these cu\-mu\-lants with the scaling result $%
\langle \!\langle g^{n }\rangle \!\rangle \sim\langle {g }\rangle ^{2-n}$
obtained \cite{AKL:86} for phase coherent samples with $L\lesssim L_{\varphi
}$ (like 1dA) leads to an apparent contradiction: not only do $\langle
\!\langle g^{n}\rangle \!\rangle $ not decrease with $n$ but $\langle
\!\langle g^{3}\rangle \!\rangle \gg \langle \!\langle g^{2}\rangle
\!\rangle $ in long samples 1dC, 1dD. A contradiction with the central
limiting theorem is even stronger: for such long wires (with $%
L\!=\!NL_{\varphi }$, $N\approx 5$ as quoted in \cite{MW})
magnetofingerprint $\delta g(B)$ arises from independent fluctuations $%
\delta G_{i}(B)$ in $N$ phase-coherent conductors connected in series, so
that the main contribution to $\langle \!\langle g^{n }\rangle \!\rangle $
is of order $\langle \!\langle G^{n }\rangle \!\rangle /N^{2n-1}$. For
sample 1dC, this would need $\langle \!\langle G^{3 }\rangle \!\rangle >
10^{2}$ and $\langle \!\langle G^{4 }\rangle \!\rangle > 10^{4}$,
inconceivable after comparing to the data for a short wire, 1dA.

\squeezetable
\begin{table}[h]
\begin{tabular}{|l|l|l|l|l|l|l|}
\multicolumn{7}{c}{Table I. \textbf{Cumulants cited in Ref.~\cite{MW}.}} \\
\hline
& $\langle {g }\rangle $ & $\langle \!\langle g^{2}\rangle \!\rangle $ & $%
\langle \!\langle g^{3}\rangle \!\rangle $ & skewness & $\langle \!\langle
g^{4}\rangle \!\rangle $ & kurtosis \\ \hline
1dA & 372.39 & 0.33 & -0.015$\pm $0.035 &  & -0.29$\pm $0.14 &  \\ \hline
1dC & \phantom{3}10.75 & 0.55$\cdot $10$^{-3}$ & \phantom{-}0.164$\pm $0.025
&  & -0.27$\pm $0.10 &  \\ \hline
1dD & \phantom{10}8.85 & 1.71$\cdot $10$^{-3}$ & \phantom{-}0.087$\pm $0.025
&  & -0.06$\pm $0.10 &  \\ \hline
\multicolumn{7}{c}{} \\[-8pt]
\multicolumn{7}{c}{Table II. \textbf{Cumulants extracted from Figs.~1b and 2
of Ref.~\cite{MW}.}} \\[2pt] \hline
1dA & 372.4 & 0.34 & -0.73$\cdot $10$^{-2}$ & -0.036 & -0.043 & -0.36 \\
\hline
1dC & \phantom{3}10.75 & 0.54$\cdot $10$^{-3}$ & \phantom{-}0.21$\cdot $10$%
^{-5}$ & \phantom{-}0.165 & -0.86$\cdot $10$^{-7}$ & -0.30 \\ \hline
1dD & \phantom{10}8.86 & 1.66$\cdot $10$^{-3}$ & \phantom{-}0.35$\cdot $10$%
^{-5}$ & \phantom{-}0.051 & -0.16$\cdot $10$^{-6}$ & -0.06 \\ \hline
\end{tabular}%
\end{table}

2.\ {\bfseries\itshape {Status of the reported statistical data.}}
We have digitized the distribution functions for three samples
from the postscript files for Figs.\ 1 and 2 of Ref.~\cite{MW} and
evaluated $\langle \!\langle g^{n}\rangle \!\rangle $ for $n\leq
4$. Following \cite{MW}, we have excluded the tail affected by the
weak anti-localization indicated   by arrows in
Fig.~2 of \cite{MW}.
Our results are displayed in Table II. While $\langle {g}\rangle $ and $%
\langle \!\langle g^{2}\rangle \!\rangle $ agree with Ref.~\cite{MW}, the
values of $\langle \!\langle g^{3}\rangle \!\rangle $ and $\langle \!\langle
g^{4}\rangle \!\rangle $ are smaller 
by 4 to 7 orders of magnitude \cite{Remark}. 
We have
also evaluated the skewness $\langle \!\langle g^{3}\rangle \!\rangle
/\langle \!\langle g^{2}\rangle \!\rangle ^{3/2}$ and the kurtosis, $\langle
\!\langle g^{4}\rangle \!\rangle /\langle \!\langle g^{2}\rangle \!\rangle
^{2}$ \cite{StatMethods} and found them to be of the order of $\langle
\!\langle g^{3}\rangle \!\rangle $ and $\langle \!\langle g^{4}\rangle
\!\rangle $ in Table I. Thus, we speculate that the data in Table I are an
artefact of a mere confusion in Ref.  \cite{MW}: 
$\langle \!\langle g^{4}\rangle
\!\rangle $ is referred to on page 146601-2 as
\textquotedblleft the fourth cumulant $\langle gggg\rangle $ or
kurtosis\textquotedblright .

3.\ {\bfseries%
\itshape {Limited applicability of the ergodicity
hypothesis (EH).}} Some of the values of $\langle \!\langle g^{3,4}\rangle
\!\rangle $ of Table II remain large in comparison with the theory. We argue
that the source of this discrepancy is not the violation of scaling but the
limited applicability of EH (the assumption that one magnetofingerprint is
sufficient to represent statistics of sample-to-sample conductance
fluctuations).

The cumulants discussed in Ref.\ \cite{MW} are sample-specific quantities $%
\langle \!\langle {g^{n}}\rangle \!\rangle {\lower4pt%
\hbox{$
\scriptstyle{\!B} $}}$, averaged over the magnetic field, $B$. For instance,
$\langle \!\langle {g^{3}}\rangle \!\rangle {\lower4pt%
\hbox{$
\scriptstyle{\!B} $}}=\langle {g^{3}}\rangle {\lower4pt%
\hbox{$
\scriptstyle{\!B} $}}-3\langle {g^{2}}\rangle {\lower4pt%
\hbox{$
\scriptstyle{\!B} $}}\langle {g}\rangle {\lower4pt%
\hbox{$ \scriptstyle{\!B}
$}}+2\langle {g}\rangle {\lower4pt\hbox{$ \scriptstyle{\!B} $}}^{\!\!\!{3}}$
and $\langle \!\langle {g^{4}}\rangle \!\rangle {\lower4pt%
\hbox{$
\scriptstyle{\!B} $}}=\langle {g^{4}}\rangle {\lower4pt%
\hbox{$
\scriptstyle{\!B} $}}-4\langle {g^{3}}\rangle {\lower4pt%
\hbox{$
\scriptstyle{\!B} $}}\langle {g}\rangle {\lower4pt%
\hbox{$ \scriptstyle{\!B}
$}}+12\langle {g^{2}}\rangle {\lower4pt\hbox{$ \scriptstyle{\!B} $}}\langle {%
g}\rangle {\lower4pt\hbox{$ \scriptstyle{\!B} $}}^{\!\!\!{2}}-3\langle {g^{2}%
}\rangle {\lower4pt\hbox{$ \scriptstyle{\!B} $}}^{\!\!\!{2}}-6\langle {g^{{}}%
}\rangle {\lower4pt\hbox{$ \scriptstyle{\!B} $}}^{\!\!\!{4}}$, \ where $%
\langle {X}\rangle _{B}=\int_{B_{1}}^{B_{1}+B_{0}}X(B)dB/B_{0}$. We have
estimated the r.m.s.\ value of these sample-to-sample fluctuations from the
average $\left\langle \langle \!\langle {g^{n}}\rangle \!\rangle {\lower4pt%
\hbox{$ \scriptstyle{\!B} $}}^{\!\!\!{2}}\right\rangle $ evaluated using the
standard combinatorial rules \cite{TAFL}: $\left\langle \langle \!\langle {%
g^{n}}\rangle \!\rangle {\lower4pt\hbox{$ \scriptstyle{\!B} $}}^{\!\!\!{2}%
}\right\rangle =\langle \!\langle g^{n}\rangle \!\rangle ^{2}+\left( {B_{c}}%
/B_{0}\right) \langle \!\langle g^{2}\rangle \!\rangle ^{n}\,n!\int_{-\infty
}^{\infty }\left[ K(x)\right] ^{n}dx+\mathcal{O}\left( {B_{c}^{2}}/{B_{0}^{2}%
}\right) $. Here $B_{c}$ and $K(x)$ are defined via the conductance
covariance as $\langle {\!\langle {g(B)g(B^{\prime })}\rangle \!}\rangle
=\langle \!\langle g^{2}\rangle \!\rangle K{([B\!-\!B^{\prime }]}/{B_{c})}$.
For a quasi-1D diffusive wire of width $w$ and length $L\gg L_{\varphi
}\gtrsim \sqrt{hD/T}>w$, $K(x)=\left[ 1+x^{2}\right] ^{-1/2}$ and $B_{c}\sim
\Phi _{0}/wL_{\varphi }$ \cite{K-ref}, so that
\begin{equation}
\left\langle \langle \!\langle {g^{n}}\rangle \!\rangle {\lower4pt%
\hbox{$
\scriptstyle{\!B} $}}^{\!\!\!{2}}\right\rangle =\left( \alpha _{n}B_{c}/{%
B_{0}}\right) \langle \!\langle g^{2}\rangle \!\rangle ^{n}+\langle
\!\langle g^{n}\rangle \!\rangle ^{2},  \label{Var}
\end{equation}%
where $\alpha _{3}\!=\!12,\ \alpha _{4}\!=\!12\pi $. Although the limit %
\mbox{$B_{0} \to \infty $} indeed yields the ergodicity $\langle \!\langle {%
g^{n}}\rangle \!\rangle {\lower4pt\hbox{$ \scriptstyle{\!B} $}}=\langle
\!\langle g^{n}\rangle \!\rangle $, even for $B_{0}=15$T and $B_{c}\sim 0.04$%
T as in \cite{MW}, the measured $\langle \!\langle {g^{n>2}}\rangle
\!\rangle {\lower4pt\hbox{$ \scriptstyle{\!B} $}}$ is dominated by the ${n}/{%
2}$ power of the variance: $\langle \!\langle {g^{3}}\rangle \!\rangle {%
\lower4pt\hbox{$ \scriptstyle{\!B} $}}\sim \pm 0.17\langle \!\langle
g^{2}\rangle \!\rangle ^{3/2}$ and $\langle \!\langle {g^{4}}\rangle
\!\rangle {\lower4pt\hbox{$ \scriptstyle{\!B} $}}\sim \pm 0.3\langle
\!\langle g^{2}\rangle \!\rangle ^{2}$. Numerically, this is comparable to
the values in Table II. Therefore, when extracted from a small number of
measured magnetofingerprints, only those values of the cumulants of
conductance distributions exceeding the first term in the r.h.s.\ of Eq.\
(1) would signal a deviation from the one-parameter scaling. This is not the
case in Ref.~\cite{MW}.

\vspace{4mm}

\noindent V.I.\ Fal'ko$^{1}$, I.V.\ Lerner$^{2}$, O.\ Tsyplyatyev$^{1}$,
I.L.\ Aleiner$^{3}$

{\small $^{1}$ Physics Department, Lancaster University, LA1 4YB, UK}

{\small $^{2}$ School of Physics, Birmingham University, B15 2TT, UK }

{\small $^{3}$ Physics Department, Columbia University, NY 10027}


\bigskip \vspace{-0.3in}

\begin{references}

\bibitem{MW} P. Mohanty and R.A. Webb, Phys. Rev. Lett. 88, 146601 (2002).

\bibitem{AKL:86}  B.L. Altshuler, V.E. Kravtsov, and I.V. Lerner,
Zh.\ Eksp.\ Teor.\ Fiz.\ 91, 2276 (1986).

\bibitem{StatMethods} {\it Elementary Statistical Methods}, G.B. Wetherill,
Methuen \& Co Ltd, London 1967.

\bibitem{Remark} A visual inspection of the
data for sample 1dC in Fig.~2 of \cite{MW} shows   that this range
of conductance values obeys $|g\!-\!\av g|<0.2$. As the $3^{rd}$
cumulant can be represented by $\gN3=\av{g\!-\!\av g}^3$,  it
cannot exceed $0.008$ while its value in Table I is $20 $ times
bigger than such an upper bound.

\bibitem{TAFL} O. Tsyplyatyev {\em et. al.}
Phys. Rev. B 68, 121301 (2003).


\bibitem{K-ref} B.L. Altshuler and D.E. Khmelnitskii, JETP Lett. 42, 359 (1986);
P.A. Lee, A.D. Stone, and H. Fukuyama, Phys. Rev. B 35, 1039 (1987).
\end{references}
\end{document}